\documentclass[%
 aip,
 amsmath,amssymb,
 reprint,%
]{revtex4-1}

\usepackage{graphicx}  % required for figures
\usepackage{dcolumn}   % required for some tables
\usepackage{bm}        % for math
\usepackage{verbatim}  % for math
\usepackage{color}
\usepackage{mathtools}
\usepackage[english]{babel}

\usepackage{graphicx}% Include figure files
\usepackage{dcolumn}% Align table columns on decimal point
\usepackage{bm}% bold math
\usepackage[utf8]{inputenc}
\usepackage[T1]{fontenc}
\usepackage{mathptmx}

\usepackage{xcolor}

\graphicspath{{figs/}}

\newcommand{\spc}{\vspace{11pt}}

\begin{document}

\title{Tuning the stability of electrochemical interfaces by electron transfer reactions}

% Authors
\author{Dimitrios Fraggedakis}
\thanks{email: dimfraged@gmail.com}
\affiliation{
   Department of Chemical Engineering,
   Massachusetts Institute of Technology,
   Cambridge, MA 02139 USA
   }
\author{Martin Z. Bazant}
\thanks{Corresponding author – email: bazant@mit.edu}
\affiliation{
   Departments of Chemical Engineering and Mathematics,
   Massachusetts Institute of Technology,
   Cambridge, MA 02139 USA
   }

\date{\today}% It is always \today, today,

\spc

\begin{abstract}
The morphology of interfaces is known to play fundamental role on the efficiency of energy–related applications, such light harvesting or ion intercalation. Altering the morphology on demand, however, is a very difficult task. Here, we show ways the morphology of interfaces can be tuned by driven electron transfer reactions. By using non-equilibrium thermodynamic stability theory, we uncover the operating conditions that alter the interfacial morphology. We apply the theory to ion intercalation and surface growth where electrochemical reactions are described using Butler-Volmer or coupled ion-electron transfer kinetics. The latter connects microscopic/quantum mechanical concepts with the morphology of electrochemical interfaces. Finally, we construct non-equilibrium phase diagrams in terms of the applied driving force (current/voltage) and discuss the importance of engineering the density of states of the electron donor in applications related to energy harvesting and storage, electrocatalysis and photocatalysis. 
\end{abstract}

\maketitle
\section{Introduction}

Pattern-forming electrochemical reactions at electrode interfaces~\cite{markovic2013electrocatalysis,stamenkovic2017energy} play a central role in many technologically relevant processes, such as electrodeposition~\cite{schlesinger2011modern,low2006electrodeposition,han2014over,han2016dendrite}, {metal battery cycling~\cite{lu2014stable,tikekar2016design,bai2016transition}, corrosion and de-alloying~\cite{mccue2016dealloying,Erlebacher2001,erlebacher2003pattern},} ion intercalation~\cite{whittingham1976electrical,Nitta2015,Li2018,lim2016origin}, electrochemical ion pumping~\cite{vayenas1992non,Ormerod2003,lu2018electrochemically}, and  {resistive switching~\cite{mazumder2012memristors,valov2013nanobatteries,gonzalez2020lithium}.} The efficiency of each process is highly dependent on several factors, one of which is the morphology of the electrochemical interface. However, the structure of the interface varies throughout the process, and this change depends on the operational conditions, as well as on the interaction of the interface with its environment.

Very recently, it has been shown that the thermodynamic stability of a driven, open system is controlled by non-equilibrium phenomena, and in particular by driven reactions~\cite{bazant2017thermodynamic}. For example, a thermodynamically stable reactive interface can become unstable, and consequently separate into multiple phases that lead to spatial inhomogeneities~\cite{Mikhailov2009,zhao2019population}. The change in the stability of a driven reactive system is directly related to the solo-autocatalytic/inhibitory nature of the reaction.

When electrochemical reactions are involved in engineering applications, it is common practice to use the phenomenological Butler-Volmer (BV) model~\cite{bard2001,newman2004}. However, when at least one electron transfer (ET) step is involved BV is not sufficient on capturing the essential physics of the reaction mechanism~\cite{kuznetsov_book}. A more detailed description is provided by ET theories, which have initially been offered by Marcus~\cite{marcus1964chemical,marcus1993} and followed by others~\cite{hush1961adiabatic,levich1966present,chidsey1991,fedorov2014ionic}. ET theories connect the reaction kinetics with microscopic/quantum mechanical material properties, e.g density of states of electron donor, which explicitly enter in the mathematical framework of the model. Therefore, for ET based reactions we are able to have a better understanding of the microscopic physics of the process, and thus realize its limitations.

\begin{figure*}[!ht]
    \centering
    \hspace{0.08in}\includegraphics[width=0.8\textwidth]{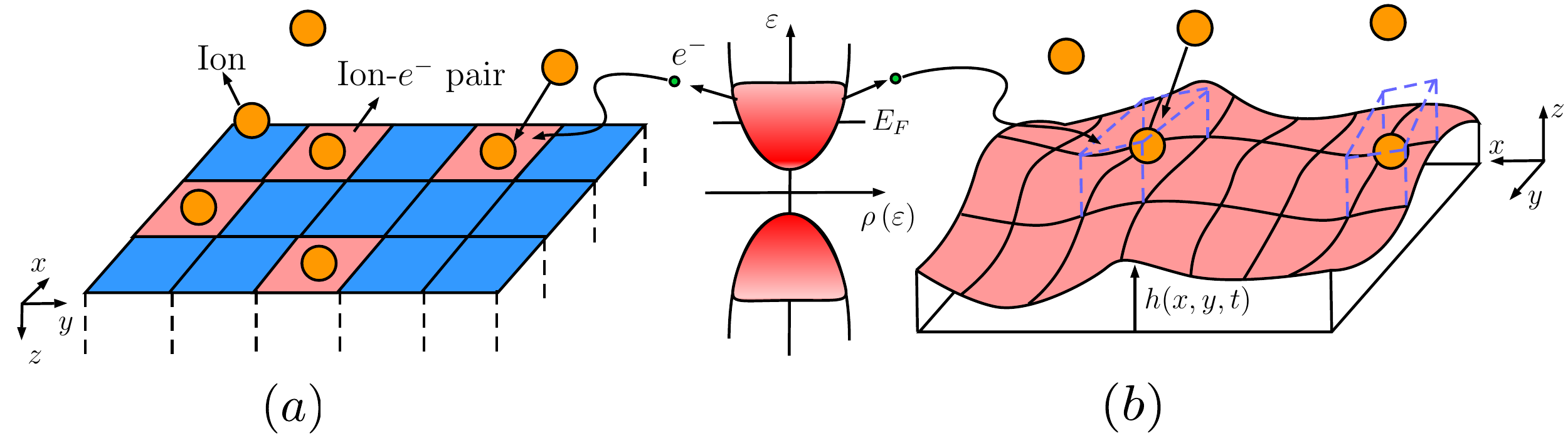}
    \caption{Two examples showing the evolution of electrochemical interfaces driven by chemical reactions. (a) The first example corresponds to ion intercalation, where ions and electrons are inserted in concerted way at a free spot on the solid interface (blue cells). Because of the attractive interactions of the ion-$e^-$ pairs with each other, the interface species might undergo a phase separation~\cite{bai2011suppression,lim2016origin} (b) The second example is related to film growth, where again, ions are `adsorbed' at a reduced spot on the existing substrate of height $h$. Curvature effects~\cite{mullins1957theory} lead to inhomogeneous film growth, which finally leads to island formation during electrodeposition~\cite{horstmann2013}.}
    \label{fig:interc_growth}
\end{figure*}

The main goal of this study is to explore the stability of evolving electrochemical interfaces that are driven by electrochemical reactions. The interfaces are open-driven systems, therefore we follow the analysis of~\cite{bazant2017thermodynamic} for different driving forces. Moreover, we are interested in understanding the impact of different electrochemical reaction models on the interfacial morphology. To do so, we focus on two different but fundamental processes, where the morphology of the interface is known to significantly affect the efficiency of practical applications. The first is ion intercalation, fig.~\ref{fig:interc_growth}(a), which is important in several technologies, \textit{viz.} Li-ion batteries~\cite{Nitta2015}, electrochromic windows~\cite{mortimer2011electrochromic,Yang2016}, neuromorphic computing devices~\cite{fuller2017li,gonzalez2020lithium}. We are also interested in film electrodeposition, fig.~\ref{fig:interc_growth}(b), which is used for the growth of catalyst nanoparticles~\cite{Lykhach2016,marks2016nanoparticle}, quantum dot formation~\cite{Penner2000}, thin-film semiconductor manufacturing~\cite{ohtsu2004method,lokhande1989electrodeposition}, formation of light-absorbing surfaces~\cite{Mandal2017}, and lithium-oxygen batteries, ~\cite{horstmann2013,Gallant2012,Mitchell2013}.  We focus on the poorly understood role of  rate-limiting interfacial transport and reactions, and neglect situations of bulk-transport-limited interfacial pattern formation~\cite{bai2016transition}, whose stability can be controlled in other ways~\cite{khoo2019linear,lu2014stable,han2016dendrite}. 

In the aforementioned cases, the interface is exposed to the environment where solvated ions are residing there, fig.~\ref{fig:interc_growth}. The ions are inserted in the system by electrochemically reducing the interface, while the electrons which participate in the reaction come from the environment. The energy level of the electrons depend on the electronic structure of the electron donor, information that is included in its density of states. Through the examination of different ET theories, we want to stress the impact of the electron donor on the interfacial stability, and how by engineer the density of states we are able to tune the topology of electrochemical interfaces.

% Chapter 2 - 
\section{Theory}
In the present work, we analyze the evolution of interfacial morphology controlled by electrochemical reactions, described by the phenomenological Butler-Volmer equation~\cite{newman2004}, electron-transfer (ET) theories~\cite{kuznetsov_book}, or coupled ion-electron transfer (CIET) theory~\cite{fraggedakis_MIET2018}. Herein, we briefly describe the main idea behind the general reaction rate theory, explain the differences between different Faradaic reaction rate models, and demonstrate under which conditions they apply based on the application of interest. Additionally, we want to control the morphology of interfaces where Faradaic reactions take place. To do so, we follow the general analysis from Ref.~\cite{bazant2017thermodynamic} and make use of non-equilibrium thermodynamic stability criteria for constructing phase diagrams in terms of the operational conditions for ion intercalation~\cite{bai2011suppression} and thin film growth~\cite{horstmann2013}. 

\subsection{Reaction Kinetics}\label{sec:2Reaction_Kinetics}
For the simple reduction reaction of cations $O^+$
$$
    O^+ + e^- \leftrightarrows R
$$
one expresses the net reaction rate $R_t$ as the difference between the forward $R_\rightarrow$ and the backward $R_\leftarrow$ elementary rates. These are described by the general theory of non-equilibrium thermodynamics, \cite{christov2012collision,keizer2012statistical,bazant2013theory}, which is based on Transition State (TS) theory~\cite{truhlar1983current}. The energy of the participating species $i$ as well as the TS energy are described by their chemical potential $\mu_i$.

Generally, when electrochemical reactions (Faradaic or not) are considered it is common practice to apply Butler-Volmer kinetics~\cite{bard2001,newman2004,butler1924part3,erdey-gruz1930theory} to describe the current $I$ as a function of the thermodynamic driving force, e.g. overpotential $\eta$. As has been shown in \cite{bazant2013theory,keizer2012statistical}, the model is rigorously derived by assuming the TS barrier $\mu_\ddag$ to be the weighted average of the reactant and product standard electrochemical potentials, where the weights are directly related to the charge transfer coefficient $\alpha$. However, the BV model is purely phenomenological and thus its parameters are not directly related to material properties~\cite{kuznetsov_book}. 

\begin{figure}[!ht]
    \centering
    \hspace{0.08in}\includegraphics[width=0.5\textwidth]{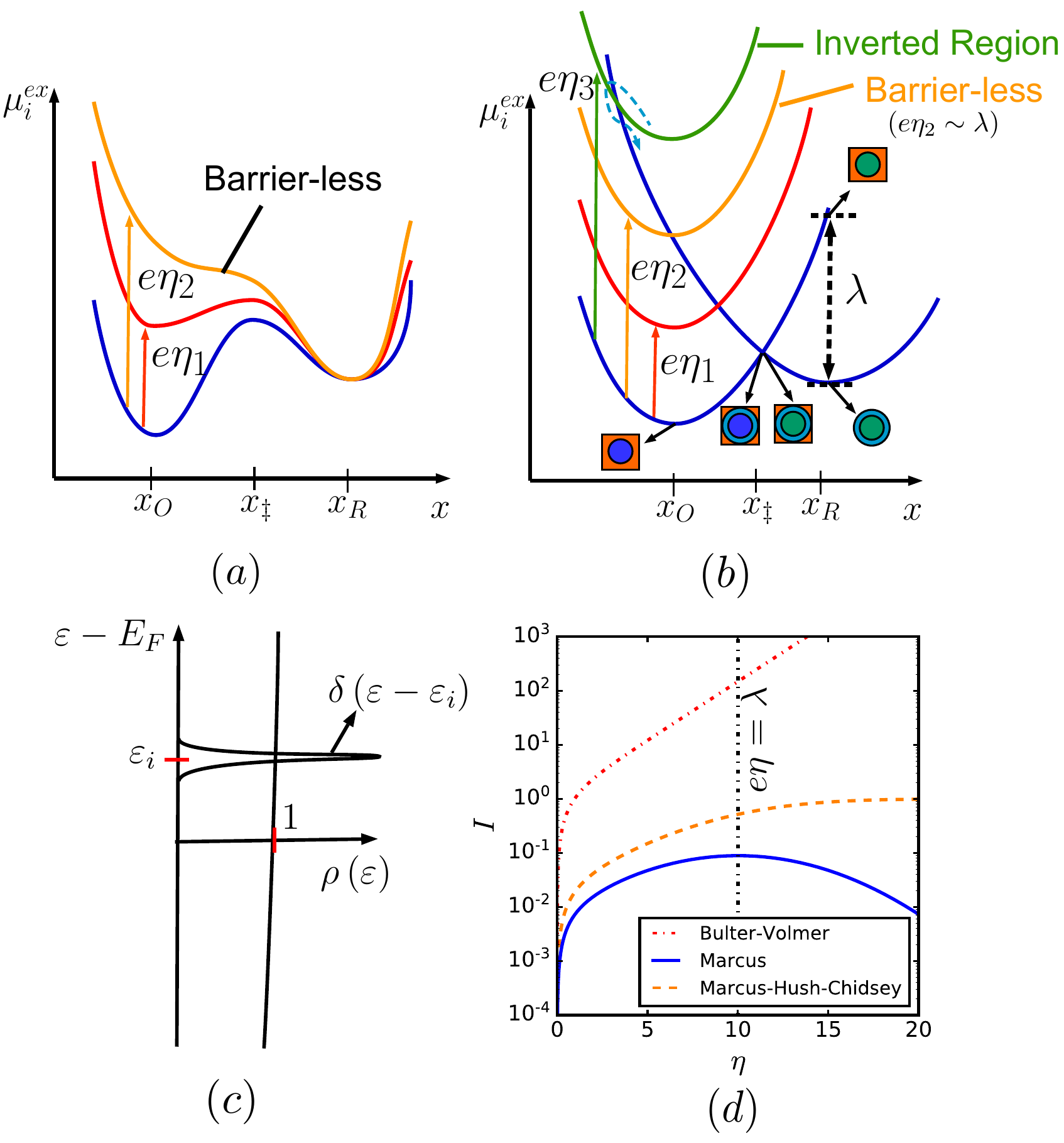}
    \caption{ Excess energy landscape of reactant ($O$)- product ($R$) species as a function of the reaction coordinate for (a) Butler-Volmer kinetics (b) electron transfer. The reaction coordinate points $x=x_O$ and $x=x_R$ correspond to the equilibrium points of the reactants (oxidized) and products (reduced) species, respectively. In both models the transition state barrier is located at $x=x_\ddag$. (c) Examples of the electron donor density of states considered in this work. (d) Current $I$ vs. applied overpotential $\eta$ for Butler-Volmer kinetics (red dotted curve), Marcus kinetics (blue curve), and Marcus-Hush-Chidsey model (orange dashed curve).}
    \label{fig:models_dos}
\end{figure}

Fig.~\ref{fig:models_dos}(a) shows the reaction energy landscape for BV kinetics in terms of an arbitrary reaction coordinate, e.g. distance of reacting species from electrode surface. In all cases, $x_O$ and $x_R$ denote the equilibrium coordinates of the reactant and product states, respectively, while $x_\ddag$ corresponds to the TS point. At equilibrium, thermal fluctuations need to provide sufficient energy to the reactants for transforming them into products. By applying a small driving force $\eta_1$ in the system (red curve), both the energy of the reactants and that of the TS point increase, but overall the absolute difference between those two decreases. Therefore, the external bias favors the reduction reaction. If larger overpotential is applied, i.e. $\eta_2$, the energy state of the reactants overcomes the TS barrier, leading to an effectively barrier-less reaction (orange curve). That said, in the BV picture there is a critical value of the applied driving force above which the resulting net reaction rate $R_t$ (or current $I = eR_t$) increases indefinitely. However, for several bulk or Faradaic reactions this trend does not apply. The current is either known to reach a limiting value with increasing $\eta$ or to show a non-monotonic behavior, where at first the current increases and after a critical $\eta$ it decreases. The region after which $I$ decreases with increasing $\eta$ is called the Marcus inverted region and is associated with the electron transfer event~\cite{marcus1965electron,marcus1993,kuznetsov_book,SchmicklerText}.

Electron transfer reactions are described by the theory introduced by Marcus, and further developed by Hush, Dogonadze, Kuznetsov, Levich and others~\cite{marcus1956oxidationreduction,marcus1965electron,levich1966present,levich1963osnovnie,kuznetsov_book,hush1961adiabatic,schmickler1986,SchmicklerText}. The main mechanism of ET involves the interactions of the electron $e^-$ participating in the reaction with the environment of the molecule/atom (solvent molecules or crystal atoms) that is reduced/oxidized. Fig.~\ref{fig:models_dos}(b) demonstrates the excess energy landscape for an ET reaction as a function of the reaction coordinate. Under this picture, the electrons are considered to be `localized' either in the reactant or product state (two-state system), while the environment undergoes thermal fluctuations. The parabolas shown represent the degree of polarization of the solvent environment. A mechanistic picture of an ET reaction is the following. Consider the equilibrium case (blue curve) for $x=x_O$, where the reactants (dark blue particle) have a solvation shell of a particular structure (orange square). Thermal fluctuations may provide enough energy to the environment helping the reactants to reach $x=x_\ddag$. There, the solvation shell has a structure which combines that of the reactants and the products one (orange square and bluish circle). Once this happens, the electron reduces the reactants species and the ET event is successful. The same mechanism is true for the reverse reaction (oxidization). The ET concept is very similar to polaron hopping, which was introduced by Landau~\cite{landau1933electron}, and further developed by Pekar, Fröhlich, Holstein~\cite{pekar1948quantum,frohlich1950xx,holstein1959studies_part1,holstein1959studies_part2,toyozawa1954theory} at around the same time as Marcus published his first paper on bulk ET.

A fundamental concept of ET theories is the reorganization energy $\lambda$~\cite{SchmicklerText,kuznetsov_book,bazant2013theory}. Its physical interpretation is understood via the following example. Consider the products excess energy landscape at equilibrium $x=x_R$ where the products (green circle) have a specific solvation shell structure (bluish green). The reorganization energy is defined as the energy required to transform the solvation shell of the products to that of the reactants without an electron transfer to take place (green circle with orange square). The same definition is given from the reactants perspective, where a different value for $\lambda$ may be defined. In the present work, we are going to limit ourselves to the same reorganization energy for both reactants and products.

As discussed earlier, the magnitude of the applied bias $\eta$ affects the behavior of the resulting current $I$. For low overpotential (red curve), the energy of the reactants increases leading to a decrease in the TS barrier, fig.~\ref{fig:models_dos}(b). As in BV model, there is a critical $\eta$ (orange curve) where the TS barrier becomes zero and the reaction becomes barrier-less. However, further increase in the applied bias (green curve) does not lead to zero activation energy, rather it starts increasing it again ($\Delta \mu^{ex} = \mu^{ex}_\ddag - \mu^{ex}_O>0$). This phenomenon leads to the celebrated Marcus inverted region where $I$ decreases with increasing $\eta$.

In bulk ET theory, the electron participating in the reaction has a single energy level $\varepsilon$. In most electrochemical systems, though, a reaction occurs nearby an electrode which supplies electrons that occupies a spectrum of energy levels (Faradaic reaction). Therefore, the density of states (DOS) of the electron donor plays a crucial role in defining the dynamics of electron transfer, as well as the final structure of the interface on which the Faradaic reaction takes place. To determine the overall reaction rate, one has to integrate over all the available energy electron levels, resulting in~\cite{kuznetsov_book,SchmicklerText,zeng2014simple}
\begin{equation}\label{eq:R_tot}
    R_t=\int\rho(\varepsilon)\left(R_\rightarrow(\varepsilon)-R_\leftarrow(\varepsilon)\right)d\varepsilon
\end{equation}

Herein, we are interested on how different DOS models affect the stability of electrochemical interfaces. In particular, we focus on the differences between a localized electron state and that of a delocalized one. In the former case, the DOS is $\rho(\varepsilon) = \delta(\varepsilon-\varepsilon_i)$, which leads to the so-called Marcus model, while in the second one, the DOS of a metallic donor is considered. To a good approximation, the metallic DOS is described by $\rho(\varepsilon) = 1$ recovering the Marcus-Hush-Chidsey (MHC) model~\cite{chidsey1991}. {In this case, when the driving force (overpotential) becomes larger than the reorganization energy $\lambda$, the celebrated `inverted-region' is lost because electrons from multiple energy states contribute to the total reaction rate, leading to what is known as reaction-limited current~\cite{kuznetsov_book,fraggedakis_MIET2018}}. Of course, different electron donor DOS can be used, e.g. for semi-conductors, semi-metals, etc.~\cite{Schmickler2017}, though the final conclusion on the stability of electrochemical interfaces does not change.

Fig.~\ref{fig:models_dos}(d) illustrates the Tafel plot for the three models considered, namely the BV, Marcus, and MHC constitutive relations. For BV, it is well-known that $I$ increases indefinitely with increasing $\eta$. For the other two cases, though, when $\eta\ge\lambda$ the current either reaches its maximum value and then decreases (Marcus model) or it attains a limiting value $I\left(\eta\rightarrow\infty\right)\simeq I_{max}$ (MHC model).

Up to this point, we described extensively the physics of different Faradaic reaction models which contain only one electron transfer as the rate limiting step. Although we have not discussed about the formulation of coupled ion-electron transfer (CIET) kinetics, the main idea is similar to the classical ET picture. CIET is based on the concerted transfer of both an ion and an electron, where the ion transfer is being described by classical transition state theory while that of the electron is based on Marcus kinetics. More details about the mathematical derivation of the model is found in~\cite{fraggedakis_MIET2018}, where the theory is shown to describe quantitatively the insertion of Li ions in FePO$_4$ (FP)~\cite{lim2016origin,Li_nat_2018}.

\subsection{Electrochemical Stability}

n
Very recently, it was shown that this is not always the case~\cite{bazant2017thermodynamic}, and the thermodynamic stability of a solution is controlled by using non-equilibrium driving forces. When the reaction rate depends on the concentration of the participating species, in addition to their chemical potentials, solo-autocatalytic/inhibitory effects participate on determining the homogeneity of the solution. Therefore, when a chemical reaction is auto-inhibited, that means with increasing products concentration the reaction rate decreases, there is a critical value in the driving force after which the system becomes thermodynamically stable. A characteristic example of such behavior is the lithiation of Li$_x$FePO$_4$~\cite{bai2011suppression,cogswell2012coherency,li2014,lim2016origin,Nadkarni2018,Li_nat_2018,nadkarni2019modeling,gonzalez2020lithium} in which the net reaction rate is a decreasing function of the concentration~\cite{fraggedakis_MIET2018,lim2016origin}.

\subsubsection{Basic Formalism of linear stability}

Herein, our main focus is to analyze the simple reaction described in Sec.\ref{sec:2Reaction_Kinetics}. Also, we consider the reactants to be drained from a reservoir, while the system contains only the products. The reaction is driven either under constant rate $R_t$ or under constant reservoir chemical potential $\mu_{res}$. In electrochemical systems these conditions are translated into constant current $I$ and constant electrode voltage $V$, respectively. 

{

We are interested in understanding the linear stability of the order parameter $c$, which can  represent either the local species concentration or the height of a domain. The general evolution equation for $c$ under reaction-limited growth is~\cite{bazant2013theory}
\begin{equation}\label{eq:order_parameter}
  \frac{\partial c}{\partial t}=R
\end{equation}
where $R$ is the reaction rate. In general, $R$ is expressed as a function of $c$, its chemical potential $\mu$, and in open systems the reservoir chemical potential $\mu_{res}$. 

For phase separating systems, the following form of the free energy $G$ is assumed
\begin{equation}\label{eq:gradient_expansion}
  G[c,\nabla c] = \int_V \left(g_h\left(c\right)+\frac{1}{2}\kappa \lvert\nabla c\rvert^2 \right)dV
\end{equation}
where $g_h(c)$ is the homogeneous energy and $\kappa$ is the penalty gradient term, which is related to the energy required to form an interface in the bulk solution~\cite{Cahn1958,Rowlinson1979}. Hence, the chemical potential $\mu$ is defined as the variational derivative of $G$
\begin{equation}\label{eq:chem_potential}
  \mu = \frac{\delta G}{\delta c} = \frac{\partial g_h}{\partial c} - \kappa \nabla^2c
\end{equation}

In this work, we examine the stability of an initially homogeneous solution with mean order parameter $\bar{c}$. Under infinitesimal perturbations, the concentration profile is of the form 
$$
  c({\mathbf{x}},{t})=\bar{c}+\varepsilon e^{\sigma t-i\mathbf{k}\cdot\mathbf{x}} 
$$
where $\sigma$ corresponds to the growth rate of the perturbation and $\mathbf{k}$ to its corresponding wave vector. Substituting $c(\mathbf{x},t)$ in eq.~\ref{eq:order_parameter} and keeping the first order terms in $\epsilon$ only, the following equation for the growth rate results
\begin{equation}\label{eq:growth_rate_sonal}
  {\sigma} =
  \frac{\partial {R}}{\partial c}+\frac{\partial {R}}{\partial {\mu}}\frac{\delta {\mu}}{\delta c}
\end{equation}
and substituting $\frac{\delta {\mu}}{\delta c}=\frac{\partial {\mu}_h}{\partial c}+{\kappa} k^2$ one arrives at~\cite{bazant2017thermodynamic}
\begin{equation}
  {\sigma} = \frac{\partial {R}}{\partial c}+\left(\frac{\partial {\mu}_h}{\partial c}+{\kappa} {k}^2\right)
  \frac{\partial {R}}{\partial {\mu}}
\end{equation}
}

{Following the analysis in~\cite{bazant2017thermodynamic} for the wavelength $k$ that maximized the growth rate $\sigma$,} the stability window is determined by solving the following equation for the critical operation conditions  
\begin{equation}\label{eq:growth}
    \widetilde{\sigma}\left(c,\widetilde{I}_c,\widetilde{V}_c\right) = \frac{\partial \widetilde{I}}{\partial c}+\frac{\partial \widetilde{I}}{\partial \widetilde{\mu}}\frac{\partial \widetilde{\mu}}{\partial c} = \widetilde{\tau}^{-1}
\end{equation}
where $\left(\widetilde{I}_c,\widetilde{V}_c\right)$ correspond to the critical value of the external bias, and $\tau$ is the characteristic time required to complete the process, e.g. the time required to fully intercalate the system~\cite{bai2011suppression} or the time to deposit an atomic layer~\cite{horstmann2013}. All quantities with $\,\,\widetilde{}$\,\, are dimensionless. Also, $\widetilde{\sigma}$ is known as the dispersion relation~\cite{bazant2017thermodynamic}. {The detailed derivation of eq.~\ref{eq:growth} when solid diffusion is considered is given in the Appendix.} 

According to classical linear stability theory, when $\widetilde{\sigma}>0$ the process is unstable and it diverges from its base state, while for $\widetilde{\sigma}<0$ all the applied perturbations decay in time and stability is preserved. In the context of ion intercalation, instability means the separation of the ionic solution in two regions, one `rich' and one `poor' in ions. For film growth, though, $\widetilde{\sigma}>0$ translates either into the film growth with increased surface roughness~\cite{kardar1986dynamic} or the formation of localized islands~\cite{horstmann2013} in the nanoscale~\cite{khoo2019linear}. Eq.~\ref{eq:growth} shows that different reaction rate mechanisms will result in different stability behavior. Therefore, by understanding how the microscopic physics and the material properties alter the phase diagram, we will be able to operate the process of interest under optimal conditions. Additionally, the theory can be used to provide engineering guidelines on the materials selection and device design to achieve the desired results.  

{

\subsection{\label{sec:models}Thermodynamics and Reaction Models for Ion intercalation and Film Growth}
\subsubsection{\label{sec:models_intercalation}Ion Intercalation}
In the case of ion intercalation~\cite{bai2011suppression}, it is common practice to use the regular solution model to describe phase-separating materials. The form of the homogeneous free energy $g_h(c)$ reads
\begin{equation}\tag{A.11}\label{eq:A.11}
  g_h(c) = \Omega c\left(1-c\right)+k_BT\left(c\ln c+(1-c)\ln (1-c)\right)
\end{equation}
where $\Omega$ is the species attraction energy, $c$ is the concentration of the inserted species and $1-c$ is the concentration of the vacancies. This model corresponds to a lattice gas. The only free parameter in eq.~\ref{eq:A.11} is $\Omega/k_BT$ which, in this study, we set $\Omega/k_BT=4.0$ corresponding to LiFePO$_4$ at room temperature~\cite{bai2011suppression}.

Here, three different reaction models are considered. These are the coupled ion-electron transfer for a metallic and a localized density of states and the Butler-Volmer model. The general reaction rate for the case of coupled ion-electron transfer is
\begin{widetext}
\begin{equation}\label{eq:A.12}
% \begin{split}
  R = \int_{-\infty}^{\infty} \rho\left(\varepsilon\right)k_0\left(n\left(\varepsilon\right)(1-c)e^{-\frac{\left(\lambda+\eta_f-\varepsilon\right)^2}{4\lambda}}-(1-n\left(\varepsilon\right) )c e^{-\frac{\left(\lambda-\eta_f-\varepsilon\right)^2}{4\lambda}}\right)d\varepsilon
% \end{split}
\end{equation}
\end{widetext}
where $n\left(\varepsilon\right)$ is the Fermi-Dirac distribution and $\eta_f = \mu-\mu_{res}-\ln c/(1-c)$. Unless otherwise specified, we set $\lambda/k_BT=8.3$~\cite{bai2014}. 

\subsubsection{\label{sec:models_epitaxi}Film Growth}
For the film growth, we adopt a thermodynamics model that is commonly used in epitaxial growth, and has been shown to describe qualitatively the formation of Li$_2$O$_2$ in Li-air batteries~\cite{horstmann2013}. The model for the homogeneous Gibbs free energy is
\begin{equation}\label{eq:A.13}
  g_h(h) = \frac{2e}{d_\bot\pi}\left[-E_0\pi h +E_1\sin^2\left(\pi h\right)-E_2e^{-\beta h^2/2}\right]
\end{equation}
where $h$ is the height of the film. The physical meaning of the parameters $E_0$, $E_1$, $E_2$ and $\beta$, as well as their values, are described in~\cite{horstmann2013}. For film growth, the ET-based reaction model expression is
\begin{widetext}
\begin{equation}\label{eq:A.14}
% \begin{split}
  R = \int_{-\infty}^{\infty} \rho\left(\varepsilon\right)k_0\left(n\left(\varepsilon\right)e^{-\frac{\left(\lambda+\eta_f-\varepsilon\right)^2}{4\lambda}}-(1-n\left(\varepsilon\right) )e^{-\frac{\left(\lambda-\eta_f-\varepsilon\right)^2}{4\lambda}}\right)d\varepsilon
% \end{split}
\end{equation}
\end{widetext}
where $n\left(\varepsilon\right)$ is the Fermi-Dirac distribution and $\eta_f = \mu(h)-\mu_{res}$.

}
% Chapter 3 - 
\section{Results}

\begin{figure*}[!ht]
    \centering
    \hspace{0.08in}\includegraphics[width=1.0\textwidth]{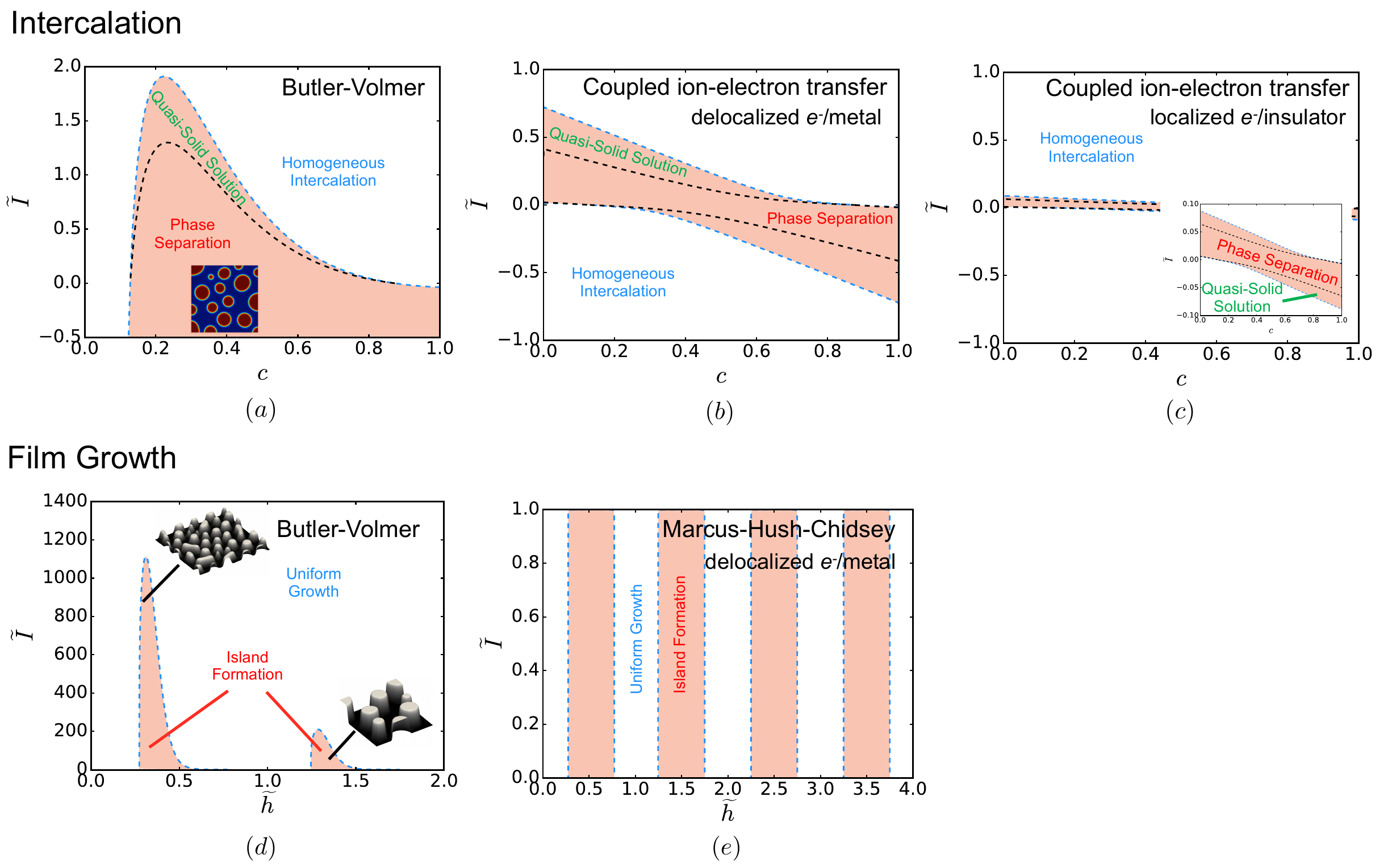}
    \caption{Phase stability diagrams for ion intercalation (a)-(c), and film growth (d)-(e), under applied constant current $\left(\widetilde{I}\right)$ conditions for different electrochemical reaction models. The light red region corresponds to the linearly unstable region, where phase separation is predicted to occur. The marginal stability curve is denoted with the light-blue dashed line. For ion intercalation, the black dashed line is used to describe the boundary of the quasi-solid solution, the region in which the applied perturbation does not have enough time to evolve until the process ends. The results for Butler-Volmer kinetics are shown in (a) and (d), while the results for the ET models are illustrated in (b),(c) and (e). For demonstration purpose snapshots of the system state at specific $c/\widetilde{h}$ are shown for the Butler-Volmer model. ET-based models produce qualitatively similar results.}
    \label{fig:const_current}
\end{figure*}

Our main scope is to show the differences on the predicted stability diagrams for different electrochemical reaction models, e.g. the phenomenological BV kinetics and the electron transfer theories. The comparison is performed in terms of ion intercalation and thin film growth. 

Fig.~\ref{fig:const_current} illustrates the non-equilibrium phase diagrams in terms of the dimensionless imposed current, and the state of the system (ion concentration $c$ in ion intercalation, film height $\widetilde{h}$ in film growth). At first sight, the differences in the predictions between BV and the ET models are apparent. Not only they differ quantitatively, but the regions which are linearly unstable (light red color) have qualitatively different bounds.

This behavior is explained in physical grounds by examining the predictions for each process separately. Irrespective of the reaction mechanism ion intercalation is known to be a solo-autoinhibitory process, $\partial \widetilde{I}/\partial c<0$ ~\cite{bazant2017thermodynamic,lim2016origin,bai2011suppression}. Therefore, a different mechanism is required to affect the second term in eq.~\ref{eq:growth} which couples the thermodynamic stability via $\partial_c\widetilde{\mu}$ with the changes of $\widetilde{I}$ as a function of $\widetilde{\mu}$. For BV kinetics, it is easily shown that $\partial \widetilde{I}/\partial\widetilde{\mu} = -\partial \widetilde{I}_\leftarrow/\partial\widetilde{\mu} \sim -e^{\widetilde{\mu}}$. On the other hand, for Marcus/MHC models the sign of $\partial \widetilde{I}/\partial\widetilde{\mu}$ is not definite, and it changes with different values of the thermodynamics driving force $\widetilde{\eta}=\widetilde{V}+\widetilde{\mu}$. Therefore, the exact reaction mechanism, and thus its mathematical form, is crucial to understanding the thermodynamic stability of a far-from equilibrium system. Having this in mind, it is not surprising why different phase diagrams for BV, Marcus and MHC kinetics are produced. In particular, for BV, phase separation is suppressed solely due to the inhibitory nature of the reaction. On the contrary, both CIET models involve not only the contribution of ions at the TS but also the effects of the thermodynamic state of the system on the electron transfer mechanism via $\widetilde{\mu}$~\cite{fraggedakis_MIET2018}. 

Returning to figs.~\ref{fig:const_current}(a)-(c), it is apparent that all models predict suppression of phase separation for $\widetilde{I}>\widetilde{I}_c(c)>0$. Here $\widetilde{I}_c$ corresponds to the critical current (light dashed line) obtained after solving $\widetilde{\sigma}\left(c,\widetilde{I}_c,\widetilde{V}_c\right)=0$. The stability boundary differs between BV and CIET models as a result of the convoluted effects of ion and electron transfer phenomena. More specifically, those effects are shown in fig.~\ref{fig:models_dos}(d) where for the same overpotential BV predicts the largest value of $\widetilde{I}=\widetilde{I}_{BV}$, Marcus model has the lowest value on $\widetilde{I}=\widetilde{I}_{M}$, while MHC predictions ($\widetilde{I}=\widetilde{I}_{MHC}$) are in between the two other models $\widetilde{I}_{BV}>\widetilde{I}_{MHC}>\widetilde{I}_{M}$.

The main differences between the models predictions occur for de-intercalation. Systems following BV kinetics are predicted to always be unstable when $\widetilde{I}<0$, as the process becomes solo-autocatalytic in the reverse direction~\cite{bazant2017thermodynamic}. This is not the case for ET models where an extended thermodynamically stable region is present. The \textit{in-situ} experiments performed by Lim et al.~\cite{lim2016origin} on the delithiation of Li$_x$FePO$_4$ showed that under moderate to large charging rates, LFP particles did not undergo phase separation, an observation that cannot be explained using BV kinetics~\cite{bazant2017thermodynamic,Li_nat_2018}. Combining this result with the predicted stability diagram for the CIET-MHC model which shows a thermodynamically stable region upon charging, fig.~\ref{fig:const_current}(b), we are able to say that LFP is a coupled ion-electron transfer limited material~\cite{fraggedakis_MIET2018}. Now, comparing the results between the two ET models, we conclude that under smaller values of the applied current, electrons with localized energy levels tend to stabilize even further the system, fig.~\ref{fig:const_current}(c).

For film growth the predicted non-equilibrium phase diagrams are very different to each other, fig.~\ref{fig:const_current}(d)-(e). For conciseness we show only the results for one ET model (MHC) as their predictions are qualitatively similar. Following Horstmann et al.~\cite{horstmann2013}, $\widetilde{I}_c(\widetilde{h})$ is predicted from $\widetilde{\sigma}\left(\widetilde{h},\widetilde{I}_c,\widetilde{V}_c\right)=\widetilde{I}$. In this case, the transition state does not include ionic effects such as excluded volume~\cite{bai2011suppression}. That translates into $\partial \widetilde{I}/\partial \widetilde{h}=0$ which makes the stability to be solely determined by balancing the second term of eq.~\ref{eq:growth} with the characteristic time $\widetilde{\tau}^{-1}=\widetilde{I}$. Here $\widetilde{\tau}$ is the time required to fully deposit an atomic layer of height $\Delta \widetilde{h}=\widetilde{h}_{i+1}-\widetilde{h}_i=1$. 

Clearly, under constant $\widetilde{I}$ the non-equilibrium stability diagram produced using ET models coincide with the equilibrium thermodynamics one, fig.~\ref{fig:const_current}(e). There are two reasons for this behavior. At first, the changes in the film chemical potential with $\widetilde{h}$ are abrupt, leading to large values of $\partial \widetilde{\mu}/\partial \widetilde{h}$. Additionally, the process attains current values which are limited by electron transfer, fig.~\ref{fig:models_dos}(d), resulting in $\left\lvert{\partial \widetilde{\mu}}/{\partial \widetilde{c}}\right\rvert \gg \widetilde{I}\left({\partial \widetilde{I}}/{\partial \widetilde{\mu}}\right)^{-1}$ for all $\widetilde{h}$ and $\widetilde{I}$. 

\begin{figure*}[!ht]
    \centering
    \hspace{0.08in}\includegraphics[width=1.0\textwidth]{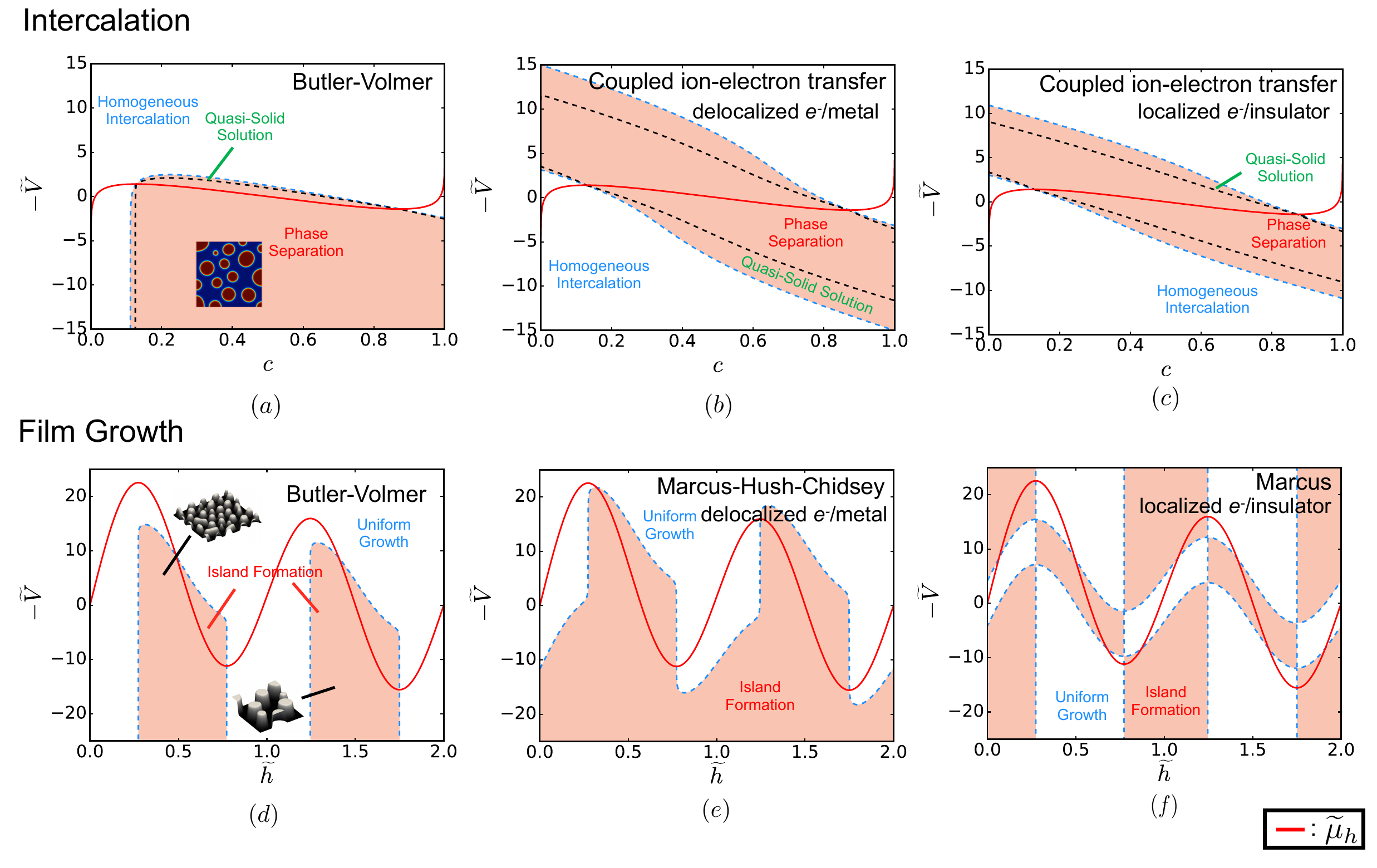}
    \caption{Phase stability diagrams for ion intercalation (a)-(c), and film growth (d)-(f), under applied constant voltage $\left(\widetilde{V}\right)$ conditions for different electrochemical reaction models. The light red region corresponds to the linearly unstable region, where phase separation is predicted to occur. The red line corresponds to the system homogeneous chemical potential. The marginal stability curve is denoted with the light-blue dashed line. For ion intercalation, the black dashed line is used to describe the boundary of the quasi-solid solution, the region in which the applied perturbation does not have enough time to evolve until the process ends. The results for Butler-Volmer kinetics are shown in (a) and (d), while the results for the ET models are illustrated in (b),(c),(e) and (f). For demonstration purpose snapshots of the system state at specific $c/\widetilde{h}$ are shown for the Butler-Volmer model. ET-based models produce qualitatively similar results.}
    \label{fig:const_volt}
\end{figure*}

For BV kinetics, the same thermodynamic stability argument is true, but the model predicts indefinite large values for $\widetilde{I}$ which is not realistic. This is the main reason Horstmann et al.~\cite{horstmann2013} were able to predict suppression of inhomogeneous Li-O$_2$ film growth, fig.~\ref{fig:const_current}(d). However, the predictions of fig.~\ref{fig:const_current}(d) do not imply much regarding the reaction mechanism of Li-O$_2$ formation. The linear stability results correspond to the initial stages of interfacial instability. As in the case of LFP, large enough current lead to a quasi-homogeneous film profile because there is not enough time for the instability to grow~\cite{bai2011suppression}. In fact, as shown in fig. 1(c) of~\cite{horstmann2013} the structure of the `homogeneous' film show the existence of some Li$_2$-O$_2$ microstructure on the CNT surfaces~\cite{Gallant2012,Mitchell2013}. Thus, it is possible Li$_2$-O$_2$ growth to be electron transfer limited, a fact that is supported by the experimentally observed and \textit{ab-initio} predicted curved Tafel plots~\cite{Viswanathan2013}.

Not only constant $\widetilde{I}$, but also constant voltage $\widetilde{V}$ stabilize an unstable solution. Fig.~\ref{fig:const_volt} shows the predicted phase diagrams under non-equilibrium conditions for both ion intercalation, figs.~\ref{fig:const_volt}(a)-(c), and film growth, figs.~\ref{fig:const_volt}(d)-(f). Additionally, the chemical potential $\widetilde{\mu}$ of the products is included (red thick line) to highlight the departure from equilibrium via $\widetilde{\eta}\left(c\right)$. 

For ion insertion ($-\widetilde{V}>\widetilde{\mu}$) all models predict suppression of any instability for a wide range of $\left(-\widetilde{V},c\right)$, figs.~\ref{fig:const_volt}(a)-(c). In general, under applied voltage the overpotential varies with time $\widetilde{\eta}(c)\simeq \widetilde{V}+\widetilde{\mu}(c)$ as a result of $c\left(\widetilde{t}\right) = \int_0^{\widetilde{t}} \widetilde{I}(\widetilde{\eta})d\widetilde{t}'$. When the solution enters the spinodal region, $\widetilde{\mu}\left(c\right)$ becomes a decreasing function leading to continuously increasing $\widetilde{\eta}$, and thus to increasing $\widetilde{I}$. This behavior gives an advantage to phase separating materials for Li-ion battery applications, as under constant $\widetilde{V}$ conditions not only we stabilize their thermodynamic state but we also achieve higher (dis)charging rates. 

From figs.~\ref{fig:const_volt}(a)-(c), BV kinetics predict more stable ion insertion compared to ET models. Let us consider the case with $-\widetilde{V}=5$. For BV the solution starts and remains inside the thermodynamically stable region for all the values of $c$ along the process. This is not true for MHC and Marcus models, though, where the material is predicted to be unstable from $c=0$ to approximately $c\simeq0.6$, and there is high probability for phase separation. The reason why this happens is understood in terms of the resulting current, which as discussed earlier, changes with increasing $c$. The evaluated $\widetilde{I}$ using BV under $\widetilde{V}=-5$ is always larger than the marginal stability boundary $\widetilde{I}_c(c)$ shown in fig.~\ref{fig:const_current}(a). On the other hand, when the reaction mechanism is associated with electron transfer $\widetilde{I}$ falls into the unstable region, figs.~\ref{fig:const_current}(b)-(c), and therefore the solution tends to phase separate. 

The conclusions for de-intercalation ($-\widetilde{V}<\widetilde{\mu}$) are very similar to the case of constant current. BV is found to be unstable for almost all sets of $\left(-\widetilde{V},c\right)$, while ET models predict suppression of phase separation. Again, when the donated electron has a single energy the differential negative resistance, $\partial \widetilde{I}/\partial \widetilde{\mu}<0$, tends to stabilize the system more effectively than in the case of a continuous energy spectrum. 

When film growth is driven under constant $\widetilde{V}$, one is able to observe very interesting behavior on the resulting phase diagrams. In particular, while under constant $\widetilde{I}$ the ET stability was determined solely by thermodynamics, fig.~\ref{fig:const_current}(e), here it is found that homogeneous film growth is stable for a wide range of applied potentials $\widetilde{V}$. When only localized electrons participate in the reaction the negative differential resistance starts playing a more active role in the stabilization mechanism. In fact, as shown from fig.~\ref{fig:const_volt}(f), there is an interchange between stable and unstable regions. More specifically, with increasing driving force the current enters the Marcus inverted region altering the sign of $\partial \widetilde{I}/\partial\widetilde{\mu}$ in eq.~\ref{eq:growth}, and thus turning a thermodynamically stable growth to an unstable one~\cite{bazant2017thermodynamic}. 

The results shown in figs.~\ref{fig:const_current} \& \ref{fig:const_volt} correspond to a specific set of material parameters. Briefly, for the intercalation case large values of the reorganization energy $\lambda$ tend to induce stability under constant current conditions, as the reaction activation barrier $E_{act}\sim\lambda/4k_BT$ increases~\cite{marcus1993,bazant2013theory,fraggedakis_MIET2018}. On the contrary, for constant $V$ higher values of $\lambda$ destabilize the thermodynamics solution. 

\begin{figure*}[!ht]
    \centering
    \hspace{0.08in}\includegraphics[width=0.8\textwidth]{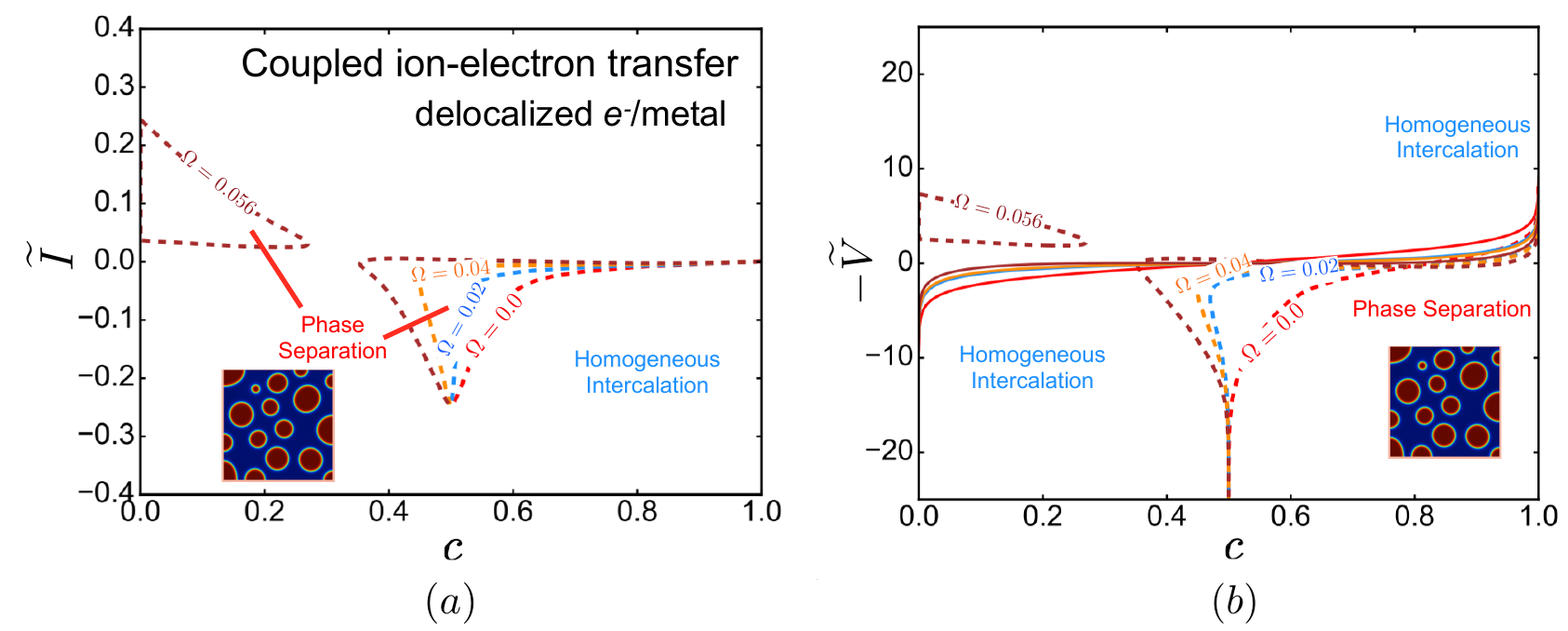}
    \caption{ 
{ 
Effect of the ion attraction energy on the non-equilibrium phase diagrams for coupled ion-electron transfer. In (a) we control applied current $\tilde{I}$, while in (b) we change the applied voltage $\tilde{V}$.
}
}
    \label{fig:omega_change}
\end{figure*}

The effects of thermodynamics factors, such as the interaction parameter $\Omega$, affect the non-equilibrium stability of the process, fig.~\ref{fig:omega_change}. In particular, with increasing attractive interactions between ions (large $\Omega$ values), the area of the unstable region increases changing the non-equilibrium phase diagram. Of interest is the case of $\Omega=0$ (no attractions) where the material is essentially in the solid-solution regime (single phase). When ions are extracted from the system ($\widetilde{I}<0$) CIET-ET models tend to destabilize the solution, inducing a non-thermodynamically favored phase separation~\cite{bazant2017thermodynamic}. This is caused by the solo-autocatalytic nature of de-intercalation and has a significant impact on the stability of several materials like those used in Li-ion batteries. A representative example is NCM which is known to be a solid solution material~\cite{yabuuchi2003novel,jouanneau2003synthesis}. However, as it is shown in \cite{Gent2016,Grenier2017,Zhou2016} for a population of cathode particles, during de-lithiation the solid solution can be destabilized leading to Li-poor and Li-rich phases. This phenomenon has been demonstrated theoretically very recently, by using population dynamics to describe the kinetics on the porous electrode scale~\cite{zhao_population_2019}. 

% Chapter 5 - 
\section{Discussion}
The results of the considered ET models stress the importance of the electron donor density of states on the morphology of electrochemical interfaces. In general, it is shown that \textit{the more localized are the electrons that participate in the electrochemical reaction, the greater the thermodynamic stability of the process}. That means, when the electrons that participate in the electrochemical reaction are initially localized, or reside in low-dimensional materials (e.g. quantum dots), then phase separation in ion intercalation or island formation in epitaxial growth is less likely to be observed. { Physically, the reason for stabilization of the interface morphology is that the more localized, poorly available electrons suppress autocatalysis and can even lead to auto-inhibitory kinetics, e.g. in the Marcus inverted region of negative differential reaction resistance~\cite{bazant2017thermodynamic}.} 

\begin{figure*}[!ht]
    \centering
    \hspace{0.08in}\includegraphics[width=1\textwidth]{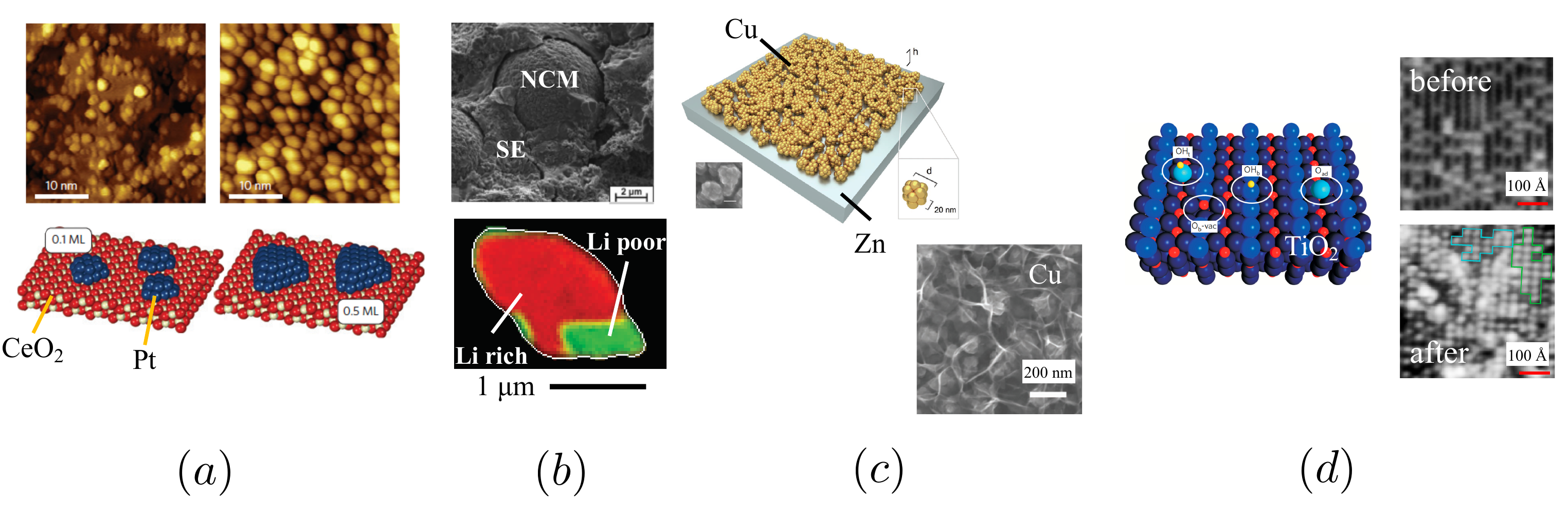}
    \caption{(a) Growth of Pt nanoparticles on CeO$_2$ substrate for applications related to catalysis. The number of electrons that are transfered during the deposition of Pt particles on CeO$_2$ control the structure of the catalytic interface. Figures are adapted from~\cite{Lykhach2016}. (b) Examples related to Li-ion intercalation in NCM and LFP. The development of stresses at interfaces between active material and solid electrolyte may cause loss of contact, and therefore induce capacity fading in the long-term operation of the battery. Figures are adapted from~\cite{Koerver2017,lim2016origin}. (c) The efficiency of light-absorbing interfaces for energy storage depends on the manufacturing conditions. More specifically, it is crucial to maximize the absorption to reflection ratio. To achieve high absorption to reflection ratio, the interfaces need not to be `rough'. Figures are adapted from~\cite{Mandal2017}. (d) Interfaces used in photocatalysis are desired to have high absorption-to-reflection ratio and large active surface area. There are cases where parasitic reactions affect the topology of the surface by forming a passivation film. We can operate the process under conditions where island formation is favorable compared to homogeneous film formation. Figures are adapted from~\cite{Hussain2017}.}
    \label{fig:exp_sys}
\end{figure*}

Our findings demonstrate new ways on tuning interfaces to have the desired topology. This is achieved by either controlling the driving force (current or voltage) of the reaction in a time-varying way, or by modifying the density of states of the electron donor, e.g. by doping. The desired topology for each application varies. For example, in (electro)catalysis it is desired to grow interfaces that have large surface area, fig.~\ref{fig:exp_sys}(a). 

In energy storage via Li-ion intercalation, we want to minimize the occurrence of phase separation, as the formation of different phases on interfaces cause the development of large elastic stresses which usually lead to fracture or delamination of the active material, fig.~\ref{fig:exp_sys}(b). Large elastic stresses between the active material and solid electrolyte may cause loss of contact, and therefore induce capacity fading in the long-term operation of the battery. This is the case in all-solid-state batteries, where inhomogeneous intercalation of Li ions leads to expansion of the host material, and consequently delamination~\cite{Koerver2017}. In commercial Li-ion batteries, the use of phase-separating materials (e.g. LiFePO$_4$, LiCoO$_2$, Li$_x$C$_6$) leads to the formation of interphases during operation. As a result, elastic stresses due to lattice mismatch develop which cause electrode particles to crack. 

Another example is energy harvesting via light absorption, where the structure of the interface exposed to light should minimize the reflection to absorption ratio, fig.~\ref{fig:exp_sys}(c). The efficiency of light-absorbing devices depends on the manufacturing conditions as the manufactured interfaces need not to be `rough'. A characteristic example is the deposition of Cu on Zn for high-efficiency solar–thermal energy conversion~\cite{Mandal2017}. The theory presented herein can be used as a guideline for the optimal selection of the manufacturing conditions, in order to produce interfaces with the desired structure. 

The same is true in photocatalytic applications, where while we desire large active surface area, the emitted light can be reflected, and thus lead to low conversions. There are cases where parasitic reactions affect the topology of the surface by forming a passivation film, fig.~\ref{fig:exp_sys}(d), which effectively decreases the active surface area and increases light reflection, respectively. While avoiding the formation of the film is difficult, we can choose the operating conditions where island formation is favorable compared to homogeneous film formation, e.g. by holding the surface voltage to levels that promote island formation.

Our model is formulated in a general way that implicitly captures system-environment interactions, e.g. double layer structure, ion (de)solvation energies, active surface area, etc., in terms of a lumped reaction constant. The present theory provides the basis to understand from first-principles the stability of interfaces undergoing electrochemical reactions, a factor which affects the performance of several applications related to energy harvesting and storage, catalysis, and electrocatalysis. 

Finally, the current status of the theory neglects elastic strain effects, which is particularly important in epitaxial growth. To include these phenomena, a more complete model is required that takes into account the formation of dislocations, as well as the spatial dependence of the developed strain upon film formation. Additionally, the formation of voids, as well as the diffusion of vacancies underneath the studied interface are expected to affect our stability results. In a future work, we will develop a general model of multilayer epitaxial growth using electrochemical reactions that takes into account the missing aspects of the present theory, and show how to exploit them to tune the morphology of interfaces. 

\section{Conclusions}
A comprehensive stability theory is presented to predict the effects of electron-transfer reaction kinetics in controlling the morphology of electrochemical interfaces The theory was applied successfully on two technologically relevant processes, those of ion intercalation and surface growth driven by electrochemical reactions. 

Using the recently developed non-equilibrium thermodynamics framework for open-driven systems~\cite{bazant2017thermodynamic}, we studied the performance of different electron transfer models on stability of a thermodynamically unstable system. In particular, we focused on the ubiquitous, but phenomenological, Butler-Volmer kinetics and on electron transfer models, which include details of the quantum/microscopic nature of the materials participating in the process (e.g. density of states of electron donor, reorganization energy of the electron environment, etc.). 

When ion intercalation is described by coupled ion-electron transfer kinetics, the process is found to be homogeneous for a larger set of the applied driving force (current $I$ or voltage $V$) with the fractional concentration $c$, compared to the BV analog. On the contrary, ET-limited surface growth is predicted to always be unstable under fixed current, leading to surfaces with increased roughness and ultimately to island formation. This is not the case when $V$ is being controlled, where ET models show a more complex behavior.

\section*{Acknowledgments}
The research was supported by the Shell International Exploration \& Production, Inc. D.F. (aka dfrag) would like to thank Neel Nadkarni, Tao Gao, Tingtao Zhou and Michael McEldrew for discussions related with the validity and application of the theory, as well as proof reading the manuscript. Finally, we want to thank the anonymous reviewers for their suggestions on the presentation of the theory.

\section*{Data Availability Statement}
The data that supports the findings of this study are available within the article.

{

\section*{Appendix}

\subsection*{\label{sec:stability_criterion}Derivation of Stability Criterion}

The system is connected to a reservoir of constant temperature $T_{rev}$, pressure $P_{res}$ and number of particles $N_{rev}$. The adsorption process is described by the following conservation law
\begin{equation}\tag{A.1}\label{eq:A.1}
  \frac{\partial c}{\partial t}=-\nabla\cdot\mathbf{j}+R
\end{equation}
where $c=N/N_{max}$ corresponds to the fractional concentration in the system, $\mathbf{j}$ is the mass flux vector and $R$ denotes the `volumetric' reaction rate. Both the flux~\cite{keizer2012statistical} and reaction rates~\cite{bazant2013theory} are formulated based on macroscopic non-equilibrium thermodynamics as follows
\begin{equation}\tag{A.2}\label{eq:A.2}
  \mathbf{j} = -M(c)c\nabla\mu
\end{equation}
\begin{equation}\tag{A.3}\label{eq:A.3}
  R = k_oe^{-\mu^{ex}_\ddag/(k_BT)}\left(e^{\mu_{1}/(k_BT)}-e^{\mu_{2}/(k_BT)}\right)
\end{equation}
In eq.{\ref{eq:A.2}}, $M(c)$ denotes the mobility of the species and $\mu$ is the chemical potential in the system. The form of eq.\ref{eq:A.3} corresponds to the rate of a general reaction of the form 
$$
  M_1=\sum_{i=1}^{N_r}s_{r,i}R_i \leftrightarrow M_2=\sum_{j=1}^{N_p}s_{p,j}P_j
$$
where $R_i$ is the reactant $i$ and $P_j$ the product $j$ along with their stoichiometric coefficients $s_{r,i}$ and $s_{p,j}$, respectively. Thus, the expressions for the chemical potentials shown in eq.\ref{eq:A.3} are $\mu_1=\sum_{i=1}^{N_r}s_{r,i}\mu_i$ and $\mu_2=\sum_{j=1}^{N_p}s_{p,j}\mu_j$, while the transition state is described by $\mu^{ex}_\ddag$~\cite{bazant2013theory}. Herein, the adsorption itself is considered as a reaction, where reservoir species $M_{res}$ are transformed into the adsorbed one $M$, leading to $\mu_1=\mu_{res}$ and $\mu_2=\mu$. It is important to note that eq.\ref{eq:A.2} is not obtained by Linear Irreversible thermodynamics (LIT) because the mobility factor is considered to depend on concentration. While, the relation may resemble that by LIT, Keizer~\cite{keizer2012statistical} showed that is derived using macroscopic non-equilibrium thermodynamics which are not constrained to cases near equilibrium~\cite{kondepudi2014modern,glansdorff1971structure}. Additionally, there may be situations where the mobility can depend on $\nabla c$~\cite{solon2016generalized}.

In order to study phase separating dynamics, the gradient expansion of the corresponding energy functional is considered to be valid. As the system of interest is held under constant temperature $T_{rev}$ and pressure $P_{res}$, the energy functional that is minimized in equilibrium is the Gibbs free energy $G$. Therefore, $G$ for the whole system is approximated as 
\begin{equation}\tag{A.4}\label{eq:A.4}
  G[c,\nabla c] = \int_V \left(g_h\left(c\right)+\frac{1}{2}\kappa \lvert\nabla c\rvert^2 \right)dV
\end{equation}
where $g_h(c)$ corresponds to the homogeneous (local equilibrium) energy landscape and $\kappa$ is the penalty gradient term, which is related to the energy required to form an interface in the bulk solution. This approximation was introduced by van der Waals~\cite{Rowlinson1979}, in the content of phase transformations and `reinvented' by Cahn \& Hilliard~\cite{Cahn1958}, for studying the spinodal decomposition of metal alloys. Hence, the chemical potential $\mu$ in eq.\ref{eq:A.2} is expressed in terms of the variational derivative of $G[c,\nabla c]$ as $\mu=\frac{\delta G}{\delta c}$. For the model given in eq.~\ref{eq:A.4}, it is found that
\begin{equation}\tag{A.5}\label{eq:A.5}
  \mu = \frac{\partial g_h}{\partial c} - \kappa \nabla^2c
\end{equation}
Substituting the flux $\mathbf{j}$ and reaction $R$ expressions in eq.\ref{eq:A.1}, the evolution equation suitable for describing  phase-separation, diffusion and reaction dynamics is obtained
\begin{equation}\tag{A.6}\label{eq:A.6}
\begin{split}
  \frac{\partial c}{\partial t}=&\nabla\cdot\left(M(c)c\nabla\left(\frac{\partial g_h}{\partial c} - \kappa \nabla^2c\right)\right) \\
  &+ k_oe^{-\mu^{ex}_\ddag/(k_BT)}\left(e^{\mu_{res}/(k_BT)}-e^{\mu/(k_BT)}\right)
\end{split}
\end{equation}
In particular the term involving $\frac{\partial g_h}{\partial c}$ describes the mechanism of diffusion, $\kappa \nabla^2c$ the formation of interfaces in the bulk and $R$ corresponds to the insertion of new particles in the system. 

It is better to express eq.~\ref{eq:A.6} in dimensionless form. To do so, all lengths are scaled with the largest length of the system $L_{ch}$, while the characteristic time is taken to be that of reaction $\tau_R=k_o^{-1}$. Also, the energy is scaled with $k_BT$. Therefore eq.~\ref{eq:A.6} reads
\begin{equation}\tag{A.7}\label{eq:A.7}
  \frac{\partial c}{\partial \tilde{t}}=\frac{1}{Da}\tilde{\nabla}\cdot\left(c(1-c)\tilde{\nabla}\left(\frac{\partial \tilde{g}_h}{\partial c} - \tilde{\kappa} 
  \tilde{\nabla}^2c\right)\right) + \tilde{R}
\end{equation}
where $Da=\frac{\tau_R}{\tau_D}=\frac{D_o}{L_{ch}^2k_o}$ corresponds to the Damkohler number and $\tau_D$ is the characteristic diffusion time $\tau_D=\frac{L_{ch}^2}{D_o}$.

In electrochemistry, both the reaction rate and the reservoir chemical potential are directly related to the current $I$ and the voltage $V$, respectively, via algebraic relations. In particular, $I=neR$, where $n$ is the number of electrons participating in the reaction and $e$ is the elementary charge, while $V=-\frac{\mu_{res}}{ne}$. Thus, the constant current condition is translated into the following constraint
\begin{equation}\tag{A.8}\label{eq:A.8}
  R_t=\int_VRdV
\end{equation}
where $R_t$ is the total current imposed on the system ~\cite{bai2011suppression,bazant2017thermodynamic}. Based on thermodynamics, it is known that once a variable is imposed on a certain value, then its conjugate one is calculated. Therefore, on the case where the total reaction rate is controlled, the reservoir chemical potential $\mu_{res}$ is the unknown variable. On the contrary, in the case of constant voltage, $\mu_{res}$ is externally controlled leading to variable reaction rate $R$. 

In the present section, the stability of an initially homogeneous solution with mean concentration $\bar{c}$ is examined. Thus, the concentration profile is considered to be of the form 
$$
  c(\tilde{\mathbf{x}},\tilde{t})=\bar{c}+\epsilon e^{\tilde{\sigma} \tilde{t}-i\tilde{\mathbf{k}}\cdot\tilde{\mathbf{x}}} 
$$
where $\tilde{\sigma}$ corresponds to the dimensionless growth rate of the perturbation and $\tilde{\mathbf{k}}$ its corresponding wave vector. Substituting $c(\tilde{\mathbf{x}},\tilde{t})$ in eq.~\ref{eq:A.7} and keeping the first order terms in $\epsilon$ only, the following equation for the growth rate results
\begin{equation}\tag{A.9}\label{eq:A.9}
  \tilde{\sigma} = -\frac{M(\bar{c})\bar{c}}{Da}\left[\frac{\partial \tilde{\mu}_h}{\partial c}+\tilde{\kappa} \tilde{k}^2\right]\tilde{k}^2+
  \frac{\partial \tilde{R}}{\partial c}+\frac{\partial \tilde{R}}{\partial \tilde{\mu}}\frac{\delta \tilde{\mu}}{\delta c}
\end{equation}
and substituting $\frac{\delta \tilde{\mu}}{\delta c}=\frac{\partial \tilde{\mu}_h}{\partial c}+\tilde{\kappa} k^2$ one arrives at~\cite{bazant2017thermodynamic}
\begin{equation}\tag{A.10}\label{eq:A.10}
  \tilde{\sigma} = \frac{\partial \tilde{R}}{\partial c}-\left(\frac{\partial \tilde{\mu}_h}{\partial c}+\tilde{\kappa} \tilde{k}^2\right)
  \left(\frac{M(\bar{c})\bar{c}}{Da}\tilde{k}^2-\frac{\partial \tilde{R}}{\partial \tilde{\mu}}\right)
\end{equation}
Solving for the maximum wavelength, $\partial_{\tilde{k}}\tilde{\sigma}=0$, it is found
$$
  \tilde{k}_{max}=\pm\left(\frac{Da\tilde{\kappa}\partial_{\tilde{\mu}}\tilde{R}-M(\bar{c})\bar{c}
  \partial_c\tilde{\mu}_h}{2\tilde{\kappa}M(\bar{c})\bar{c}}\right)^{1/2}
$$
Therefore, the critical conditions that should be applied depend on the way of interpreting the result. In particular, considering always $\tilde{\mu}_{res,c}$ to be the unknown of 
$$
  \tilde{\sigma}\left(\tilde{k}_{max},\tilde{\mu}_{res,c},\bar{c}\right)=0
$$
and by substituting $\tilde{\mu}_{res,c}$ in $\tilde{R}_c$ one finds the marginal curve for constant current.

It is important to make some comments on the form of $\tilde{k}_{max}$. It is true that the wavelength does not have imaginary values~\cite{bazant2017thermodynamic}, leading to constraints in the form of $\tilde{k}_{max}$. In particular, for $M(\bar{c})\bar{c}\partial_c\tilde{\mu}_h>Da\tilde{\kappa}\partial_{\tilde{\mu}}\tilde{R}$, it is apparent that $\tilde{k}_{max} = 0$. By definition, $\partial_{\tilde{\mu}}\tilde{R}$ is always negative, leading to the fact that in the vicinity of the spinodals $\left(\partial_c\tilde{\mu}_h\sim0\right)$ the phase separation is reaction- and not diffusion-controlled. Another interesting point is the existence of $\partial_c \tilde{R}$ in the expression for $\tilde{\sigma}$. In particular, this term contains the explicit dependence of the transition state on the concentration, and given its behavior, the stability of the system is affected~\cite{bazant2017thermodynamic}. In general, it is true that when a solution is thermodynamically unstable, then $\partial_c\tilde{\mu}_h<0$, leading to spinodal decomposition~\cite{Cahn1961}. But for auto-inhibitory reactions, where $\partial_c \tilde{R}<0$, $\tilde{\sigma}$ will remain stable even though $\partial_c\tilde{\mu}_h<0$.

}

% \nocite{*}
\bibliography{mhc_final_clean} 

\end{document}